\documentclass[12pt]{iopart}

\usepackage{iopams}  
\usepackage{graphicx}
\usepackage{siunitx,cite}
\usepackage{color}
\usepackage[font=small,format=plain,labelfont=bf,up,textfont=normal,up]{caption}

\usepackage[breaklinks=true,colorlinks=true,linkcolor=blue,urlcolor=blue,citecolor=blue]{hyperref}

\begin{document}

\title[2D simulation of quantum reflection]{Two-dimensional simulation of quantum reflection}

\author{Emanuele Galiffi}
\address{Department of Physics, Imperial College London, Prince Consort Rd, London SW7 2BB, UK}
\address{Physikalisches Institut, Universit\"at Heidelberg, Im Neuenheimer Feld 226, D-69120 Heidelberg}
\ead{emanuele.galiffi12@imperial.ac.uk}

\author{Christoph S\"underhauf}
\address{Institut f\"ur Theoretische Physik,
Philosophenweg 16, Universit\"at Heidelberg, D-69120 Heidelberg}
\ead{C.Suenderhauf@ThPhys.Uni-Heidelberg.DE}

\author{Maarten DeKieviet}
\address{Physikalisches Institut, Universit\"at Heidelberg, Im Neuenheimer Feld 226, D-69120 Heidelberg}
\ead{maarten.dekieviet@physik.uni-heidelberg.de}

\author[cor1]{Sandro Wimberger}
\address{Dipartimento di Scienze Matematiche, Fisiche ed Informatiche, Universit\`a di Parma, Parco Area delle Scienze n. 7/a, I-43124 Parma, Italy}
\address{INFN, Sezione di Milano Bicocca, Gruppo Collegato di Parma, Parma, Italy}
\address{Institut f\"ur Theoretische Physik, Philosophenweg 12, Universit\"at Heidelberg, D-69120 Heidelberg}
\eads{\mailto{sandromarcel.wimberger@unipr.it}} 

\begin{abstract}
A propagation method for the scattering of a quantum wave packet from a potential surface is presented. It is used to model the quantum reflection of single atoms from a corrugated (metallic) surface. Our numerical procedure works well in two spatial dimensions requiring only reasonable amounts of memory and computing time. The effects of the surface corrugation on the reflectivity are investigated via simulations with a paradigm potential. These indicate that our approach should allow for future tests of realistic, effective potentials obtained from theory in a quantitative comparison to experimental data. 
\end{abstract}

\vspace{2pc}

\section{Introduction}

Quantum reflection, i.e. the reflection of a quantum object in the absence of a classical turning point~\cite{carraro1998sticking, garrido2011paradoxical}, has attracted an increasing number of studies, triggered by the developments in the field of matter-wave optics for a variety of experimental platforms~\cite{zhao2011quantum,schewe2009focusing,shimizu2001specular,abele2014high,PhysRevA.53.319,PhysRevX.4.011029,schmiedmayer,DeKieviet2011,dufour2014quantum,PhysRevLett.50.990,PhysRevLett.63.1689,PhysRevLett.71.1589,PhysRevLett.93.223201,shimizu2002giant,pasquini2006low,zhao2010coherent,whittaker2015spectroscopic}. 

While the effect on itself has been widely known since the early days of quantum mechanics~\cite{lennard1936}, several papers, both theoretical~\cite{maitra1996semiclassical,PhysRevB.17.2147,echenique1976reflectivity,echenique1976scattering} and experimental~\cite{shimizu2001specular,zhao2011quantum}, were recently published that develop an insight into this purely quantum effect, which is crucial for modern surface science. Since most analytical treatments of quantum reflection  rely on approximations, exact numerical efforts are needed to verify their regime of applicability. In parallel, matter-wave experiments enable tests of Casimir-Polder interactions~\cite{friedrich2002quantum,PhysRevA.78.010902, PhysRevLett.105.133203, PhysRevA.87.012901, PhysRevLett.104.083201,buhmann2012casimir,whittaker2015spectroscopic, laliotis2014casimir,sukenik1993measurement}, that require  quantitative simulations to bridge the gap between theory and experiment. A recent study investigates the effect of a periodically driven surface in one dimension\cite{herwerth2013quantum} using a phenomenological atom-surface Casimir-Polder interaction potential. In two dimensions, a time-independent approach studies matter-wave diffraction due to quantum reflection from a doped surface~\cite{PhysRevA.91.013614}.

While time-independent approaches are computationally more efficient when the mere final reflection amplitude is needed, a time-dependent method enables the treatment of time-dependent Hamiltonians, and offers more insight into dynamical details of the scattering process. In this work we apply for the first time a two-dimensional time-dependent numerical scheme, that is suitable for the study of non-separable potentials. We use a toy model potential, based on a phenomenological generalization of the well known $1/r^3$ van der Waals potential~\cite{herwerth2013quantum}, to investigate the zero-order effects of a periodic corrugation on the reflectivity and connect to the one-dimensional (1D) results obtained in \cite{herwerth2013quantum} for reference. Our results represent a proof-of-principle that quantum reflection from realistic Casimir-Polder potentials can now be investigated quantitatively in two-dimensions (2D) via an optimized time-dependent numerical propagation procedure.

\section{Numerical Method}

For 2D dynamics, in which there is significant coupling between the two dimensions, semiclassical WKB-type methods, that work well in 1D static problems, are known to fail \cite{maitra1996semiclassical,wkb2004,friedrich2004working,galiffi2015thesis}. Numerical wave packet propagation schemes, that are based on split-operator techniques to treat the higher spatial dimensions (e.g.\cite{pisa1998,wimberger2005resonant}) are not very stable either in the case of strong non-separability.

Therefore, the numerical procedure we choose makes use of the implicit, norm-preserving Crank-Nicholson scheme with backward and forward substitution~\cite{press2007numerical}. Assuming that the Hamiltonian is constant over a small enough time interval $\Delta t$, $\hat{H}(\hat{x},\hat{y})$, the finite difference time-dependent Schr\"odinger equation in Cayley's form reads:
\begin{equation}
    (1 + \frac{i \hat{H}}{2 \hbar} \delta t) \vec{\psi}_{t+\delta t} = (1 - \frac{i \hat{H}}{2 \hbar} \delta t) \vec{\psi}_{t}
    \label{eq:TDSE_Cayley}
\end{equation}
where $\hat{H}$ is represented as an $( N_x \cdot N_y) \times (N_x \cdot N_y)$ matrix. Equation (\ref{eq:TDSE_Cayley}) needs to be solved for $\psi_{t+\delta t}$ at each time step, so that the wave function can be known at any time during the scattering process.
If three-point approximations for second derivatives are used along the $x$- and $y$-axes, the resulting structure of the Hamiltonian in the position basis is a tridiagonal block matrix with fringes~\cite{press2007numerical}: The elements of the $\vec{\psi}(x,y)$ vector are indexed $\vec{\psi}(x,y)=[\psi_{0,0}, \psi_{0,1}, ... \psi_{0,N_y}, \psi_{1,0}, ... ,\psi_{N_x-1,N_y-1}]$, where the subscripts represent the indices on the grid along the $x$- and $y$-axes respectively, and $N_x$ and $N_y$ are total numbers of grid points. It can be shown that this basis minimizes the matrix band to the width $N_y$~\cite{sunderhauf2014}.

\textcolor{black}{The Cayley form used in Eq. (\ref{eq:TDSE_Cayley}) has the advantage that a relatively large time step can be chosen without compromising the norm of the evolved state. Using a grid step which depends on position, at least in our realization with a three-point discretization, would make the corresponding matrices $\hat H$ non-symmetric, and consequently, we would lose the mentioned advantage of perfect norm preservation.}

In order to optimize the solution to Eq.(\ref{eq:TDSE_Cayley}), the matrix which \textcolor{black}{multiplies $\vec{\psi}_{t+\delta t}$} is decomposed into the product of a lower triangular factor and its transpose, via Cholesky (or $LL^t$) decomposition~\cite{press2007numerical}, so that only one triangular matrix must be stored. For time-independent Hamiltonians, the Cholesky factor, which depends on the potential parameters and the spatial and temporal step sizes, only needs to be computed and stored once, assuming that these are kept fixed. Equation (\ref{eq:TDSE_Cayley}) can thus be solved by standard forward and backward substitution~\cite{golub2012matrix}.

The band preservation property of Cholesky decomposition \cite{tsukerman2007computational} implies that the resulting triangular Cholesky factors inherit the banded structure of the Hamiltonian, so that only a limited number of matrix elements of order $N_x\times N_y^2$ needs to be saved. This reduces the memory requirements dramatically. Therefore, the memory cost scales linearly with the number of grid points along one ($x$), and quadratically along the other axis ($y$). This means that this method is most efficient for problems in which the numerical grids are very different along the two dimensions: In other words, for long and thin grids. In this method, periodic boundary conditions are easily implemented. This suggests to study systems with a periodicity along the axis, along which grid points are more expensive. In fact, it turns out that, in this basis, the additional matrix elements necessary to introduce periodic boundary conditions along the $y$-axis lie at the end of each $x$-block, within $N_y$ elements from the main diagonal. Therefore, they do not increase the size of the band, so that memory usage is unchanged and additional runtime is negligible. 

Since we are interested in a scattering problem involving a particle, we assume for it an initial Gaussian wave packet with standard deviations $\sigma_x$ along the axis normal to the 1D surface ($x$) and $\sigma_y$ along the in-plane ($y$) direction. Its average position $( \langle x \rangle, \langle y \rangle)$ and average momentum $(\langle p_x \rangle, \langle p_y \rangle)$. The wave packet initially approaches the surface with $p_x<0$. The momentum representation $\tilde{\psi}(p_x,p_y)$ is computed via a 2D Fast Fourier Transform at different timesteps~\cite{press2007numerical}, and the total reflectivity is best computed \cite{herwerth2013quantum} by integrating the latter over the entire $p_y$ axis and the positive region of the $p_x$ axis, so that
\begin{equation}
R(t) = \int_{-\infty}^{\infty}{dp_y\int_{0}^{\infty}{dp_x |\tilde{\psi}(p_x,p_y,t)}|^2}\,.
\end{equation}
accounts for all reflected components of the wave packet at time $t$. Since $R$ depends on time, its final value is taken only after it has reached the stationary regime. This may be obscured when, in addition to quantum reflection sharply oscillating spurious reflections arise from $\psi(x,y)$ hitting the $x$-boundary of our numerical grid. We eliminate these by suppressing the transmitted part of the wave function by multiplying it at each timestep by a smooth sigmoidal filter function, whose parameters are optimised to minimise such artificial reflections (see \cite{herwerth2013quantum} for details).

\section{Paradigm Potential}

In one dimension, the atom-surface interaction can be described, in the short-range v/d Waals limit as:
\begin{equation}
    V_{1D}(r) = -\frac{C_3}{(r-r_0)^3}
    \label{eq:1D_potential}
\end{equation}
where $C_3$ and $r_0$ are material coefficients and $r$ is the distance of the atom along the normal to the point-like surface~\cite{herwerth2013quantum,vidali1991potentials,ihm1987systematic}.

In order to simulate the effect of a small corrugation along the surface, we generalise equation (\ref{eq:1D_potential}) by introducing a periodic modulation in the distance to the surface\textcolor{black}{, which leads to a first order approximation to the results in \cite{1126-6708-2003-06-018, DeKieviet2011}}, such that 
\begin{equation}
r-r_0 \to r(x,y) = x-A\sin{(\frac{2\pi}{L}y + \phi)}\,, 
\label{eq:sub}
\end{equation}
where $A$, $L$ and $\phi$ are the amplitude, wavelength, and phase of the corrugation respectively. This defines our paradigm potential
\begin{equation}
    V_{2D}(x,y) = - \frac{C_3}{ ( x - A \sin{ ( \frac{2\pi}{L}y + \phi ) )^3 } } \,,
    \label{eq:para}
\end{equation}
for testing the influence of the periodic corrugation on the quantum reflectivity $R(t\to \infty)$.
The singularity, which occurs as $x\to A\sin{(\frac{2\pi}{L}y + \phi)}$, is removed by introducing an artificial cutoff length $\Delta$, beyond which the potential is continued as a parabola with vertex in $r(x,y) = 0$. For $r(x,y) < 0$, we impose $\frac{\partial V}{\partial x} = 0$, so that $V(x,y)$ becomes independent of $x$. Continuity and differentiability are imposed at the cutoff, as a generalisation of the approach introduced by~\cite{herwerth2013quantum} for the 1D case.

The complete, smooth 2D paradigm potential can thus be written as
\begin{eqnarray}
V_{2D}(x,y) = \left\{\begin{array}{lr}
    -\frac{C_3}{r^3(x,y)}, & r(x,y)\geq\Delta  \\
    \frac{3C_3}{2\Delta^5} r^2(x,y)-\frac{5C_3}{2\Delta^3}, & 0\leq r(x,y)\leq \Delta\\
    -\frac{5C_3}{2\Delta^3}, & r(x,y)\leq0
    \end{array} \right. \,,
    \label{eq:2D_pot}
\end{eqnarray}
Introducing a cutoff length $\Delta$ for the potential induces oscillations in $R$ as $\Delta$ is varied, as shown in later sections, and the effective reflectivity $\bar R$ can be computed by averaging over these oscillations \cite{herwerth2013quantum}. \textcolor{black}{We end this section by noting that this continuation is effectively 1D, as it simply shifts the point at which the parabola is matched with Eq. (\ref{eq:para}), so that the locus of points at which the potential is continued satisfies the equation $x-A\sin{(\frac{2\pi}{L}y +\phi)}=\Delta$, meaning that it is an equipotential line. Our simplified potential is invariant under a translation which leaves $r(x,y)$ in Eq. (\ref{eq:sub}) unchanged. This enables such a simple continuation to be straightforwardly implemented while preserving the 2D nature of the problem. In fact, an additional arbitrary parameter would need to be introduced in the continuation if less symmetric potentials are used. Physically, a continuation along the $x$ axis is justified by the fact that the gradient of the potential, which is the key physical parameter of this problem, has a $y$-component which is proportional to the ratio $A/L$, so that in the small amplitude regime the steepest potential gradient is always along the $x$-axis}.

\section{Results of numerical simulations}

\begin{figure}[tb!]
 \centering
 \includegraphics[width=0.95\textwidth]{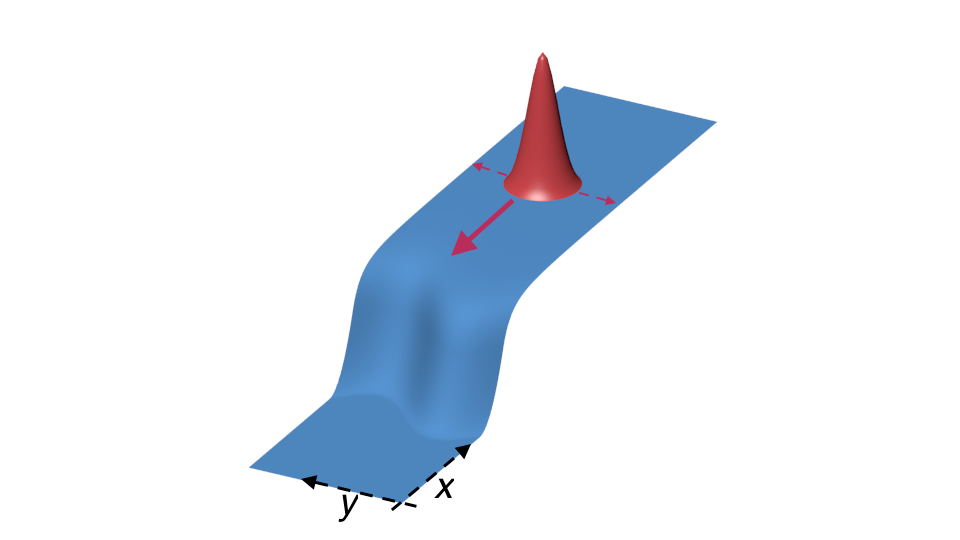}
 \caption{\label{fig:sketch_pot_psi} A sketch of the physical setup: the blue surface represents the potential landscape, whereas the squared modulus of the initial gaussian wavepacket is shown in red. The red solid arrow indicates the direction of propagation, whereas the dotted lines represent the spreading of the wavepacket during the propagation.}
\end{figure}

We now present the results of our tests on convergence and further investigations of the effects of a small corrugation on the reflectivity for the potential defined in Eq. (\ref{eq:2D_pot}). We consider a corrugation wavelength $L=$ \SI{100}{\nano\metre}, which coincides with the size of our numerical grid along the $y$-axis, for which periodic boundary conditions are used. Orthogonal to it our grid spans the region \SI{-1.5}{\micro\metre} $\leq x\leq$ \SI{5.0}{\micro\metre} along the surface normal. The interaction constant was chosen $C_3 =$ \SI{4.0e-50}{J} which corresponds to the v/d Waals potential between a \textcolor{black}{$^3$He} atom and a Au surface~\cite{ihm1987systematic, vidali1991potentials} ($m \approx 5.01\times 10^{-27}{\rm \,kg}$~\cite{wieser2013atomic}). A sketch of the potential with the initial wavepacket is shown as a 3D plot in figure \ref{fig:sketch_pot_psi}.

The departure from the flat surface case is investigated by gradually increasing the amplitude $A$ of the corrugation, which introduces a coupling between the two dimensions in the Hamiltonian.
The width of the incoming Gaussian wave packet along the normal axis $x$ is fixed at $\sigma_x = $ \SI{80}{\nano\metre}. The initial wave packet is always placed at $\langle x(0) \rangle = $ \SI{2.0}{\micro\metre}, with an initial average velocity $(\langle v_x(0)\rangle, \langle v_y(0)\rangle) = (\langle p_x(0)\rangle/m, \langle p_y(0)\rangle/m) = (2.0,0.0)$ \si{\metre\per\second}.

\subsection{Convergence}

Timestep convergence tests confirm that $dt \leq $ \SI{6.0}{\nano\second} guarantees good convergence, with relative residuals of order $10^{-4}$ for all spatial step sizes at which spatial convergence is achieved. Therefore, in the following simulations, we use $dt=$ \SI{5.0}{\nano\second}. To test our numerical machinery, we study convergence with respect to spatial grid sampling. Simulations were performed using a small corrugation $A/L=0.1$. We fix the phase $\phi$ of the corrugation to zero at the initial position of the wave packet $\langle y\rangle =0$ . The width of the initial wave packet along the $y$-axis is $\sigma_y =$ \SI{8}{\nano\metre}.

\begin{figure}[tb!]
 \centering
 \includegraphics[width=0.95\textwidth]{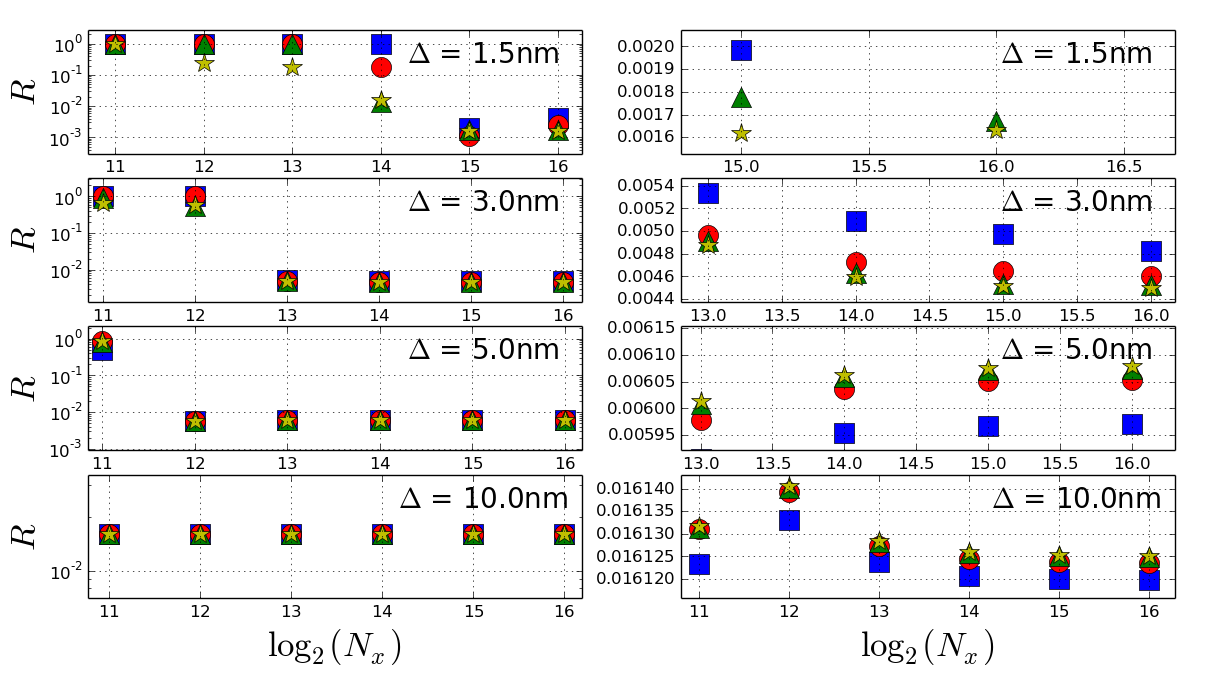}
 \caption{\label{fig:spatial_convergence} Each of the subplots shows the reflectivity $R$ vs number of grid points $N_x$ for different $N_y$ (\opensquare : $ N_y = 2^{5}$, \opencircle : $N_y = 2^{6}$, \opentriangle : $N_y = 2^7$, $\star$ : $N_y = 2^8$). \textcolor{black}{Spatial convergence in the $x$ direction is seen along the $x$ axis of the panels. Convergence along the $y$ direction is seen following the different symbols in each panel at fixed $N_x$}. Each row corresponds to a different value of the cut-off length $\Delta$. The first column shows the behaviour of $R$ in logarithmic scale, whereas the second shows a zoom into the convergence regime in linear scale. The amplitude of the corrugation is $A/L=0.1$.}
\end{figure}

The results of these tests w.r.t. cut-off distance $\Delta$ are shown in figure \ref{fig:spatial_convergence}. In each subplot, the reflectivity is plotted as a function of the number of grid points used along the normal axis $N_x$. Powers of two are used for the number of grid points in order to facilitate the use of a Fast-Fourier Transform~\cite{press2007numerical}. The different symbols represent the level of precision along the $y$-axis, $N_y$.

Each row in fig 1 corresponds to a fixed value of the cutoff length $\Delta$. The left column of plots show $R$ in logarithmic scale, whereas the right one shows a zoom into the convergence region on a linear scale. 
Figure 1 shows how smaller values of $\Delta$ require a higher spatial resolution for convergence. This is expected for the toy potential of Eqs. (\ref{eq:para}) and (\ref{eq:2D_pot}): Since the potential becomes singular as $x\to A\sin{(\omega y + \phi)}$, the large potential gradient encountered at small values of $\Delta$ results in a poor sampling of the potential landscape. As a result, a higher precision in the spatial grid is needed for convergence. The abrupt increase of $R$ as the numerics breaks down is a known effect from previous studies\cite{herwerth2013quantum,sunderhauf2014}. At that point, the potential is effectively sensed by the numerics as a step, which generally has a higher reflectivity than a soft potential~\cite{garrido2011paradoxical}. The relative errors associated to the present numerical method, which was thoroughly tested in 1D and compared to exact solutions~\cite{herwerth2013quantum}, are of the order of \textcolor{black}{$\Delta R/R \approx 1\%$}. \textcolor{black}{The aforementioned 1D treatments show that convergence with respect to $\Delta$ is achieved when $\Delta$ is in the \SI{}{\nano\metre} regime and fractions thereof, where it can be extracted efficiently via logarithmic averaging~\cite{herwerth2013quantum}. Taking this into account, we observe that good convergence can be achieved with a number of grid points along the $x$-axis $N_x\geq 2^{15}$ and along the $y$-axis $N_y \geq 2^7$ up to a cut-off of $\Delta = $ \SI{3}{\nano\metre}, whereas a finer grid enables convergence at $\Delta = $ \SI{1.5}{\nano\metre}. Finer grids can still be used, especially along the $x$ axis, in order to reduce $\Delta$ to the sub-nanometre regime.}

The scaling of the computational costs with different grid sizes is shown in tables 1a through 1c.

\begin{table}[b]
\centering
\begin{tabular}{l|lll}
\multicolumn{4}{c}{a) Memory Usage (GB)} \\ \hline\hline
$N_x$  & $2^{14}$    & $2^{15}$    & $2^{16}$     \\ \cline{1-4}
$N_y$ = $2^5$       & 0.4   & 0.8   & 1.6    \\ \cline{1-4} 
$N_y$ = $2^6$       & 1.4   & 2.4 & 4.8  \\ \cline{1-4} 
$N_y$ = $2^7$ & 5.0 & 9.0  & 18.0  \\ \cline{1-4} 
$N_y$ = $2^8$ & 18.0   & 35.0   & 69.0
\end{tabular}
\hfill
\begin{tabular}{lll}
\multicolumn{3}{c}{b) $LL^{T}$-Time ($h$)} \\ \cline{1-3}\hline\hline
$2^{14}$    & $2^{15}$   & $2^{16}$     \\ \cline{1-3} 
-   & -   & -    \\ \cline{1-3} 
-   & 0.2 & 0.3  \\ \cline{1-3} 
0.3 & 1.5  & 3  \\ \cline{1-3} 
6   & 12.5   & 23
\end{tabular}
\hfill
\begin{tabular}{lll}
\multicolumn{3}{c}{c) Time Step ($s$)} \\ \cline{1-3}\hline\hline
$2^{14}$    & $2^{15}$    & $2^{16}$     \\ \cline{1-3} 
3.6   & 4.0   & 8.25    \\ \cline{1-3} 
10   & 18 & 36  \\ \cline{1-3} 
36 & 169  & 342  \\ \cline{1-3} 
295   & 729   & 1450
\end{tabular}
\caption[justification=justified,singlelinecheck=false]{\label{tab:table_costs}The computational cost of the simulations for different numbers of grid points along the $x$ (columns) and $y$ (rows) axes. Table 1a shows the memory usage in GB, 1b the approximate number of hours needed to perform a single Cholesky factorisation ($LL^{T}$), and 1c gives an upper bound for the propagation time per single time step (dt=5 ns) in seconds. Machines used for computations requiring more than 4GB of memory feature AMD Opteron 6282 SE CPUs.}
\end{table}

\subsection{Effects of cutoff and corrugation}

\begin{figure}[tb]
\includegraphics[width=0.445\textwidth]{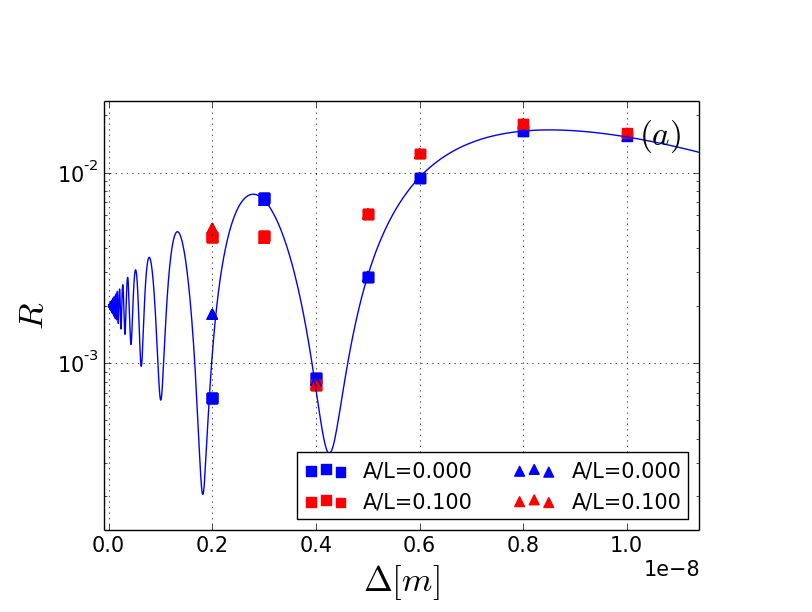}
\includegraphics[width=0.475\textwidth]{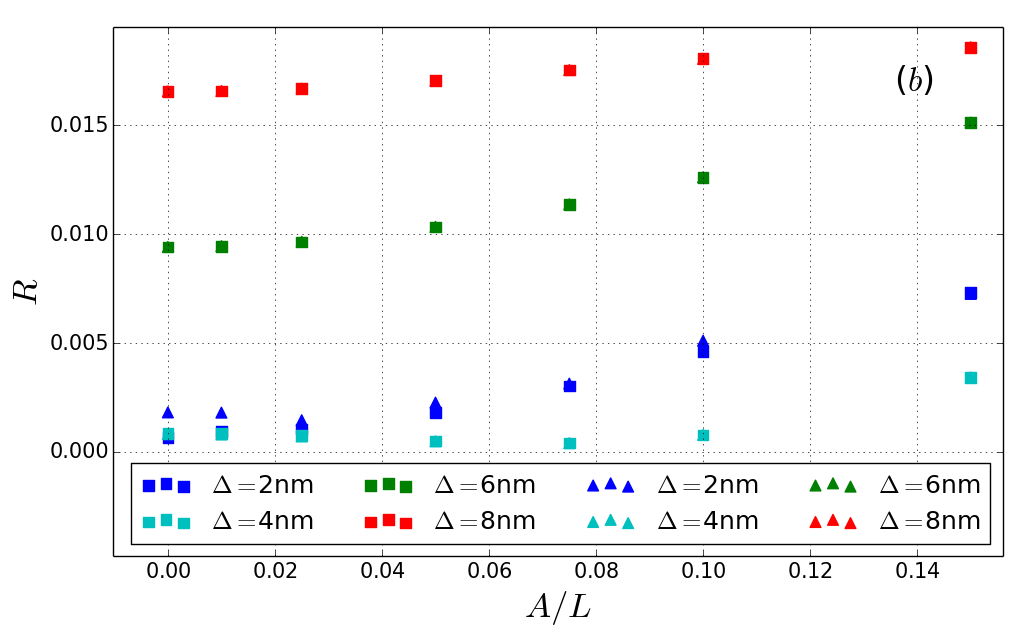}
\caption{The squares represent simulations with $N_x=2^{14}$, and triangles $N_x=2^{15}$. (a) Reflectivity as a function of cutoff length for different amplitudes $A$ at $L=10 \rm \, nm$. Where the symbols are distinguishable, convergence is worse. The continuous lines show the 1D time-independent results, which coincides with the flat surface case. (b) Reflectivity as a function of corrugation amplitude $A$ for different cutoff length $\Delta$.}
\label{fig:subplots_A_Delta}
\end{figure}

Next, we systematically study the effect of the cutoff length $\Delta$ on the reflectivity. Starting from a $A/L=0$ (i.e. the flat surface), we increase the corrugation amplitude gradually. Figure \ref{fig:subplots_A_Delta}(a) shows the oscillations in $R$ as a function of $\Delta$, which appear analogous to those in our 1D results~\cite{herwerth2013quantum}. The reflectivity for the flat surface is compared to 1D time-independent simulations (continuous line) for different cutoff lengths $\Delta$. $N_y = 2^7$ was used in all simulations. For smaller values of the cutoff length, the symbols for $N_x=2^{14}$ and $N_x=2^{15}$ can be distinguished, which signifies poor convergence. The 1D results are accurately reproduced by our 2D method in the flat surface case; we find a relative difference between the two methods of less than $\Delta R/R\leq1\%$ down to $\Delta = $ \SI{3.0}{\nano\metre}; below this cutoff convergence is lost.

We also calculate the reflectivity as a function of $A/L$ for fixed cutoff points (figure \ref{fig:subplots_A_Delta}b). We observe that $R$ changes with the corrugation amplitude $A$ for all values of the cutoff. These deviations are smaller for larger values of the cutoff. This is expected, because if the potential is parabolically continued very far from the surface, the effects of the corrugation are too small to influence the dynamics significantly.

\subsection{Effect of the width of the wave packet}

\begin{figure}[tb!]
 \centering
 \includegraphics[width=0.65\textwidth]{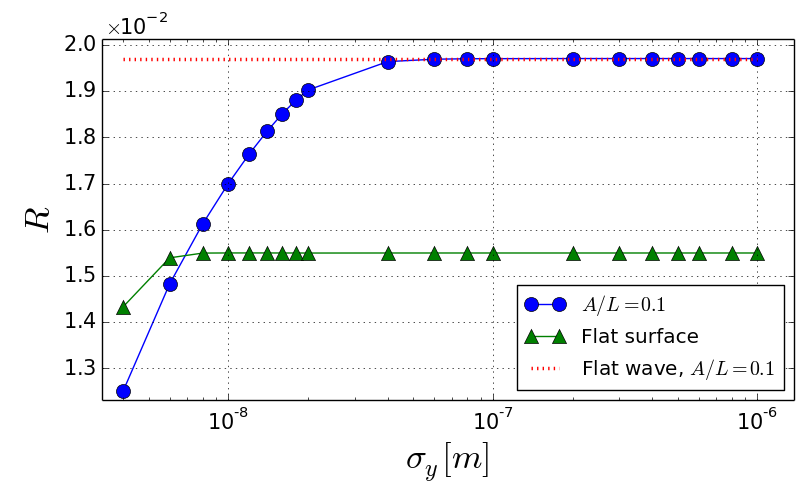}
 \caption{\label{fig:R_vs_y_width} Reflectivity as a function of the $y$-width of the initial gaussian wave packet for a flat surface (triangles) and a corrugated surface with amplitude to period ratio $A/L=0.1$ (circles) with $\Delta = $ \SI{10}{\nano\metre}. The red dotted line represents the saturation value obtained for a wave packet which is flat along the $y$-direction (i.e. $\sigma_y \gg L$).}
\end{figure}

In order to observe the effects of coupling introduced by the corrugation, we study the effect of increasing the width of the incoming wave packet $\sigma_y$ on the calculated reflectivity. \textcolor{black}{Since we are only interested in the effect of widening the initial wave, we just use the part of the Gaussian, with width $\sigma_y$ in the second ($y$) direction, which is within the considered interval $L$, with proper normalization of the $y$ marginal of the modulus of the wave function. For $\sigma_y \ll L$, much smaller than the periodicity $L$ in $y$ direction, this corresponds to a usual Gaussian, whilst in the other limit $\sigma_y \gg L$ we obtain an essentially flat profile in the $y$ direction.}
The results are shown in figure \ref{fig:R_vs_y_width} for $N_x=2^{14}$ and $N_y=2^7$. For wider wave packets, an increase in the reflectivity is observed for the corrugated surface ($A/L=0.1$), whereas essentially no dependence is observed for the flat surface. This is exactly what is expected: for the flat surface no scattering occurs in the $y$-direction, whereas the corrugation couples the $x$ and $y$ momentum components of the wave packet. Since the initial state is a Gaussian wave packet with $\langle v_y \rangle=$ \SI{0}{\metre\per\second}, a wider distribution in position space (along $y$) results in a momentum distribution more localised around $p_y=$ \SI{0}{\kilogram\metre\per\second}. Because quantum reflection decreases steeply with increasing incident momentum~\cite{herwerth2013quantum}, waves having fewer spectral components of larger momentum are reflected more. As the width of the wave packet approaches the size of the numerical grid, the reflectivity saturates, so that the incoming wave packet effectively becomes uniform along the $y$-axis. For $\sigma_y <$ \SI{4.0}{\nano\metre}, the grid sampling is not sufficient for a precise resolution of the initial wave packet, and a sharp, non-physical decrease in the reflectivity is obtained (for very small $\sigma_y$).

\section{Conclusions and Perspectives}

In this work, we have presented numerical simulations on quantum reflection, using an optimised Crank-Nicholson scheme in combination with a Cholesky decomposition tailored to a banded matrix system. We demonstrate that this 2D approach reproduces our earlier 1D results in the limit of zero corrugation. In addition, we show that this numerical method can be applied for the time-dependent simulation of quantum reflection from a realistic 2D non-separable potential. While this method is especially suitable for periodic problems such as uniaxially corrugated surfaces, it can in principle be used for non-periodic structures as well (modulo additional computational costs). As a first practical result we have shown that the coupling between the two dimensions introduces a dependence of the reflectivity on the corrugation amplitude.

While the feasibility of this type of computations has been demonstrated here for a phenomenological 2D non-separable potential, the possible parameter space is still very broad and can be explored and adjusted according to the specific needs of the experiments. We postpone such investigations to the future when more realistic 2D potentials will be available for modeling actual experiments. Our numerical technique allows us to investigate quantum reflection from any given conservative potential and can test theories that produce different effective potentials, amongst each other, and with respect to experimental results.


As shown in~\cite{herwerth2013quantum}, higher order approximations for second derivatives in the Hamiltonian enable better precision. Since the consequent extra terms along the $y$-axis do not increase the size of the band, the computational costs for their implementation should be negligible, thus obviating for the extra cost of grid points along the $y$-axis (quadratic in $N_y$). In addition, the time step $dt$ might be made adaptive by performing Cholesky factorisation at different points during the propagation, depending on the numerical precision needs.

From a numerical point of view, the use of the Demko-Moss-Smith theorem for the decay of matrix elements in the inverse of a banded matrix away from the band~\cite{demko1984decay}, may save  memory since only a smaller approximation of the inverse of the matrix in the lhs. of Eq. (\ref{eq:TDSE_Cayley}) is needed. This may allow us to replace the iterative forward/backward substitution procedure by matrix-vector products only, which are simple to implement and fast to execute. Alternatively, spectral approaches may be considered, which exploit the symmetry and structure of the Hamiltonian, such as the one proposed in~\cite{molinari2013identities}. Finally, the forward-backward sutstitution procedure might be implemented in parallel, thus reducing dramatically the time required for each propagation step.

We end this paper by discussing some practical consequences of our work: the length scale for quantum reflection ideally matches that of the periodicity and corrugation height encountered in nano-materials. For conventional atom-surface scattering long-range order and atomic flatness are required not to loose the entire beam intensity in diffuse scattering. In quantum reflection, however, the turning points lie much further away from the surface, which makes our numerical propagation method well suited even for large roughness that nano-materials typically exhibit. Here, not only their static properties, but also their dynamics can be investigated, i.e. when the Hamiltonian is explicitly time-dependent. Moreover, our method allows for studies of time-dependent details of the quantum-reflection scattering process itself that cannot be addressed directly using time-independent approaches. As an example, the propagation of surface plasmons and bulk plasmonic excitations that extend all the way to the surface are very relevant processes in organic electronic materials. These open questions can now be investigated systematically. Furthermore the dynamics within the novel magnetic materials developed in recent years can studied with quantum reflection, using scatting particles with a spin (such as $^3$He used in atomic beam spin echo spectrometers). An example of the search for structure and dynamics in micro-structured artificial magnetic assemblies is discussed in \cite{Ahrend2015292}. Note that these magnetic structures do not necessarily possess a physical corrugation that comes along with the magnetic one. The ever decreasing physical size of a single bit in magnetic storage devices and, therewith correlated, their increasing volatility, offer a true playground for the time and length scales of our numerical method. From a more fundamental point of view, our method may steer and optimize the design of a (dynamic) beam splitter for atoms that functions without laser light.

\section*{Acknowledgements}
We thank very much Stefan Buhmann and his group for discussions, and Elmar Bittner for his continuous support in using the ITP computer cluster at Heidelberg. EG acknowledges financial support from the Graduate Academy Heidelberg by the PROMOS (DAAD) program and the Physikalisches Institut Heidelberg, as well as Matthew Foulkes and Miguel Oliveira for support in the use of computing resources at Imperial College, and Sam Palmer for his help with graphics.

\section*{References}

\bibliographystyle{iopart-num}


\end{document}